\documentstyle[epsfig,12pt]{article}

\newcommand{\be}{\begin{equation}}
\newcommand{\ee}{\end{equation}}
\newcommand{\bea}{\begin{eqnarray}}
\newcommand{\eea}{\end{eqnarray}}
\newcommand{\bean}{\begin{eqnarray*}}
\newcommand{\eean}{\end{eqnarray*}}

\newcommand{\gapproxeq}{\lower
.7ex\hbox{$\;\stackrel{\textstyle >}{\sim}\;$}}
\newcommand{\lapproxeq}{\lower
.7ex\hbox{$\;\stackrel{\textstyle <}{\sim}\;$}}

\begin{document}
\begin{titlepage}
\begin{tabbing}
wwwwwwwwwwright hand corner using tabbing so it looks neat and in \= \kill
 \> {hep-ph/0102xxx}   \\
 \> {OUTP-0107P}   \\
 \> {JLAB-THY-01-01}   \\
\> {6 February 2001}
\end{tabbing}
\baselineskip=18pt
\vskip 0.5in
\begin{center}
{\bf \LARGE The origins of quark-hadron duality: How does
the square of the sum become the sum of the squares?}\\
\vspace*{0.7in}
{\large Frank E. Close}\footnote{\tt{e-mail: F.E.Close@rl.ac.uk}} \\
\vspace{.1in}
{\it Dept of Theoretical Physics}\\
{\it University of Oxford, Keble Rd., Oxford, OX1 3NP, England}\\
\vspace{0.1in}
{\large Nathan Isgur}\footnote{\tt{e-mail: isgur@jlab.org}} \\
{\it Jefferson Lab}\\
{\it 12000 Jefferson Avenue, Newport News, Virginia 23606, USA}\\
\end{center}
\begin{abstract}
Bloom-Gilman duality demonstrates empirically
that the electroproduction of $N^*$'s at low
momentum transfers averages smoothly around the scaling curve measured
at large momentum transfers. The latter is proportional to the
sum of the squares of the constituent charges whereas the former
involves the coherent excitation of resonances and is driven by the
square of summed constituent charges. We determine the minimal
necessary conditions
for this equality to be realised so that duality can occur
 and consider the implications for a range of processes that may be
studied soon at CEBAF.
\end{abstract}
\end{titlepage}
\setcounter{page}{2}
\par

When protons are probed by electron beams at energies
of tens or hundreds of GeV, and at correspondingly large
momentum transfers ($Q^2$), the scattering probability (summarised
in the structure function $F_2(W^2,Q^2)$ where $W$ is the
mass of the hadronic system) is rather simple. It exhibits the well known
property of scale invariance where $F_2 \sim F_2(W^2/Q^2)$,
 with small corrections that are
well understood from perturbative QCD. Furthermore, the magnitude
of $F_2(W^2/Q^2)$ is proportional to the sum of the squares of the
(quark and antiquark) constituent charges.

Before the advent of QCD, Bloom and Gilman discovered \cite{bloom} an
empirical property of the data, namely
 that the electroproduction of $N^*$'s at lower energies and
momentum transfers averages smoothly around the scaling curve measured
at large momentum transfers. During the subsequent three decades,
and especially following the advent of quantum chromodynamics,
this enigma has received considerable theoretical attention \cite{duality}.
The literature has primarily focussed on understanding how the $Q^2$
dependence of resonance excitation
can conspire to mimic the $x' \sim \frac{Q^2}{W^2 + Q^2}$ dependence of
the deep inelastic data.

Recent high precision
data from Jefferson Lab \cite{jlab1} have shown
that this duality is observed for both proton and neutron targets,
at least for spin averaged scattering, and that for the proton it occurs
locally, {\it i.e.}, resonance by resonance.
This has sparked a renewed interest in the origin of low energy duality,
including a recent analysis in the context of a large-$N_c$-based
relativistic quark model \cite{IJMV}. This work focussed on the dynamics
required for a confined struck quark to behave as though it were free, but remarked in
passing on the additional conditions on quark charges and dynamics required for
duality to be realized locally. These data and the work of Ref. \cite{IJMV}
have led us to focus in this paper on the latter issue,
 namely the circumstances whereby
$F_2(x')$ - whose magnitude is in proportion to the
sum of the squares of the
(quark and antiquark) constituent charges - can in general match with
 the excitation of individual resonances which is driven by the
coherently summed square of constituent charges.

 In this note
we draw attention to the necessary conditions
for this duality to occur in general. We illustrate the
essential physical principle in a simple pedagogical model of a
system of two spinless electrically charged consitutents. This is then
generalised to the more realistic case of three spinning quarks.
Implications for both spin-dependent and spin-independent structure functions
for proton and neutron targets will be displayed, local deviations
from duality are predicted, and the
possibility that there could be ``precocious" factorisation in
semi-inclusive hadroproduction is discussed. These ideas
 promise a lively program of
experimental investigation for CEBAF.

{\bf A Simple Model}

Consider a composite state made of two equal mass
scalars, ``quarks" $q_1,q_2$ with charges
$e_1,e_2 $ respectively at positions $\vec{r}_{1,2}$. The ground state
wavefunction is $\psi_0(\vec{r})$, where $\vec{r}_{1,2} = \vec{R} \pm \vec{r}/2$
defines the centre of mass and internal spatial degrees of freedom. 
A photon of momentum
$\vec{q}$ is absorbed with an amplitude proportional to
$\Sigma_i e_i exp(i \vec{q} \cdot \vec{r}_i)$,
 which excites  a ``resonant" state with
angular momentum L, described by the wavefunction $\psi_L(\vec{r})$.
 
Focussing on the internal
coordinate $ \vec{r}$, the transition amplitude is proportional to

$(e_1 + e_2) (exp(i \vec{q} \cdot \vec{r}/2) + exp(-i \vec{q} \cdot \vec{r}/2))
+(e_1 - e_2) (exp(i \vec{q} \cdot \vec{r}/2) - exp(-i \vec{q} \cdot \vec{r}/2))$

\noindent The expansion,
$exp(iqz/2) = \Sigma_L i^L P_L(cos \theta) j_L(qr/2) (2L + 1)$,
projects out the even and odd partial waves such that the amplitude is
proportional to
\begin{equation}
M \sim \int dr r^2 \psi_L^*(r) \psi_0(r) j_L(qr/2) [(e_1 + e_2) \delta_{L=even} +
(e_1 - e_2) \delta_{L=odd}].
\label{eq:r1}
\end{equation}

The resulting structure function, summed over resonance excitations, will
have the form

 \begin{equation}
F(q) \sim \sum_{n=0}^{\infty} [F_{2n}(q) (e_1 + e_2)^2 +
F_{2n+1}(q) (e_1 - e_2)^2]
\label{eq:r2}
\end{equation}

\noindent which in general will be  proportional to $e_1^2 + e_2 ^2$
only if the odd and even $L$ states sum to equal strengths. 

We now derive
the circumstances under which this can occur.
Consider the $uu $, $(ud \pm du)/\sqrt 2$ and $dd$ composite systems
subject to the following
process: $W^+ + (dd) \to (ud \pm du)/\sqrt 2 \to
W^- + (uu)$. The overall Bose symmetry of the system will constrain
the $ud + (-) du$ intermediate states to have $L = 2n$
 ($L=2n + 1$) respectively. Following analogous
steps to those above, the amplitude will have the form

\begin{equation}
A\left ( W^+(q) + (dd) \to W^-(q) + (uu)\right )
 \sim \sum_{n=0}^{\infty} [F_{2n}(q) -
F_{2n+1}(q) ]
\label{eq:r3}
\end{equation}

\noindent In this case the $t-$channel involves transfer of
exotic quantum numbers (charge 2, corresponding to constituents $uu\bar{d}\bar{d}$)
and in the absence of such
states, the amplitude must vanish according to the theory of
hadronic duality \cite{hararirosner}, and as is seen consistently in hadronic data.
 This forces
\begin{equation}
 \sum_{n=0}^{\infty} F_{2n}(q) \equiv
\sum_{n=0}^{\infty} F_{2n+1}(q)
\label{eq:r4}
\end{equation}

\noindent which with Eqn.(\ref{eq:r2}) leads to

\begin{equation}
F(q) \sim \sum_{n=0}^{\infty} F_{n}(q) (e_1^2 + e_2^2)
\label{eq:r5}
\end{equation}

\noindent whereby the square of the sum has become the sum of the
squares.

This simple example exposes the physics rather clearly. The excitation
amplitudes to resonance states contain both diagonal ($e_1^2 + e_2^2)$
 and higher twist
terms ($\pm 2 e_1 e_2$) in the flavour basis. The former set add
constructively for any $L$ and the sum over the complete set of states
can now logically give
  the
deep inelastic curve \cite{duality}; the latter enter with opposite
phases for even and odd $L$ and destructively interfere. The critical feature
that this exposes is that {\it at least one complete set of resonances of
each symmetry-type has been summed over}.

In a non-relativistic SHO model it is possible to see how the above
conspiracy arises. The contribution to $F(q)$ from the
$n (\equiv L+2k)$ set of degenerate levels ($k$ being the radial
and $L$ the orbital quantum number) is
$F_n(q) \sim (n!)^{-1}(q^2R^2)^n e^{-q^2R^2}$ from which one can immediately
see that $\Sigma_n F_n(q) = 1$. 
It is interesting to note that any individual contribution,
$F_n$, reaches its  maximum value when $q^2 R^2=n$,
 at which point $F_{n} = F_{n-1}$. 
This coincidence is true for all juxtaposed partial waves at 
their peaks, which gives a rapid approach to the equality
of $F_{odd}$ and $F_{even}$.
Analytically in the SHO one finds

\begin{equation}
F(q) \sim (e_1^2 + e_2^2) + 2e_1e_2 e^{-4q^2R^2}
\label{eq:r6}
\end{equation}

\noindent whereby the coherent (duality-violating) terms vanish
like the fourth power of the elastic form factor.

{\bf SHO Quark Model}

This pedagogical example contains all the essential physics
needed to understand the
more realistic case of the harmonic oscillator quark model. The example of spinless
constituents above
involved only electric multipoles, whereas introduction of spin
involves both electric and magnetic multipole contributions.
The symmetric and antisymmetric states generalise, respectively,
to the ${\bf 56}$ and
${\bf 70}$ representations of $SU(6)_{flavor-spin}$. The destructive
interference in the $s-$channel sum is as before and is now
correlated with the overall Fermi antisymmetry of the $SU(6) \times
SU(3)_{color} \times \psi (r)$ wavefunctions. The quark-parton model
scaling curves proportional to $\Sigma_i e_i^2$
 then obtain if  $F_1({\bf 56}) \equiv F_1({\bf 70})$ and if
the interaction of the photon with the magnetic moment of the quarks
dominates.

This numerology was demonstrated long ago in Ref. \cite{cgk},
which exhibited the decomposition
of the structure functions for $\gamma + N \to$ hadrons when the $\gamma$
and $N$ spins are antiparallel ($\sigma_{1/2}$) or parallel ($\sigma_{3/2}$)
(i.e., when the net spin projection along the initial photon direction is
 $1/2;3/2$). The relative strengths of the supermultiplets for
the case where magnetic interactions dominate are
listed in Table 1 from which $g_1^{p,n} \sim \sigma_{1/2} - \sigma_{3/2}$
and $F_1^{p,n} \sim  \sigma_{1/2} + \sigma_{3/2}$ can be constructed. The
familiar leading twist ratios  are obtained immediately from an
equal weighting of the $\bf 56$ and $\bf 70$ contributions:
$F^n/F^p = 2/3$; $A^p \equiv g_1^p/F_1^p = 5/9$; $A^n = 0$.

\vskip 0.2in
Table 1: Relative Photoproduction Strengths in the Quark Model

$
\begin{array}{lcccc}
\hline
SU(6): &\sigma^p_{1/2}  &\sigma^p_{3/2} &\sigma^n_{1/2}
 &\sigma^n_{3/2} \\
\hline
{\bf 56} & 11 & 6 & 6 & 6 \\
{\bf 70} & 10 & 0 & 3 & 3 \\
\hline
\end{array}
$

\vskip 0.2in

\noindent These results imply that for $F_1^{p,n}$ and $g_1^p$
duality will not be realised unless the $\bf 56$ and $\bf 70$ states
($N, \Delta(1236)$ and the negative parity $N^*, \Delta^*$ with $W \leq
1.8 $ GeV) have been integrated over. However, for $g_1^n$ there is
the tantalising possibility that duality may be more localised as the
$\bf 56$ alone, consisting of
$N, \Delta(1236)$ already satisfies  $\int dW g_1^n(W,q) = 0$ in this
idealised model, and the dominance of magnetic interactions is also
realised for the neutron target. In reality SU(6) is broken, in particular by the differing
masses of the $N$ and $\Delta$ due to higher twist
one-gluon-exchange hyperfine effects in QCD. Hence
it will be interesting to investigate these general ideas in explicit quark
models to predict the detailed approach to duality in a world where SU(6)
is broken. Some qualitative ideas on the $x \to 1$ dependence of
$F^n/F^p$ and $A^{n,p}$ already
exist in the
literature\cite{fc73,su6break}; with the above insights into how
the leading twist results relate to the symmetries of the excited
hadronic states, explicit calculations summing over
the above resonances and
 incorporating SU(6) breaking in QCD can now be developed and
their implications for CEBAF in particular explored.

When we break Table 1 into its $SU(3) \times SU(2)$ content, we see that for the
proton duality may be satisfied by $W \leq 1.6$ GeV. This is because the
${\bf 70;^4 8}$ and ${\bf 70;^2 10}$, which are at
$\sim 1.7$ GeV, make negligible contributions (Table 2). For neutron targets,
by contrast, one must include the ${\bf 70;^4 8}$, which necessitates integrating
up to 1.8 GeV. This region above 1.7 GeV also contains ${\bf 56}$ at $N=2$
in the harmonic oscillator. Thus we anticipate systematic deviations from
local duality, in accord with the new data from Ref. \cite{jlab1}.
These data showed that the $S_{11}(1530)$ region
(${\bf 70;^2 8}$)
and the $F_{15}(1680)$ (${\bf 56;^2 8}$) are enhanced relative
to the deep inelastic scaling curve for proton targets. As Table 2
shows, the {\bf 70} contribution for a proton target is concentrated in
the ${\bf 70^28}$, hence the enhancement seen in Ref. \cite{jlab1}.

\vskip 0.2in

Table 2: Relative Photoproduction Strengths of ${\bf 56,0^+}$ and ${\bf 70,1^-}$ Multiplets

 $
\begin{array}{lcccccc}
\hline SU(6): &{\bf [56,0^+]^2 8} & {\bf [56,0^+]^4 10} & {\bf
[70,1^-]^2 8} & {\bf [70,1^-]^4 8} & {\bf [70,1^-]^2 10}  & total
 \\
\hline
F_1^p & 9 & 8 & 9 & 0 & 1 & 27 \\
F_1^n & 4 & 8 & 1 & 4 & 1 &  18 \\
g_1^p & 9 & -4 & 9 & 0 & 1 &  15  \\
g_1^n & 4 & -4 & 1 & -2 & 1 &  0  \\
\hline
\end{array}
$

\vskip 0.3in

\noindent In contrast to the proton case, this
 table predicts that for neutron targets, the
$S_{11}(1530)$ region
(${\bf [70,1^-]^2 8}$) will fall $\bf below$ the
scaling curve. The third resonance region, containing
${\bf [70,1^-]^4 8}$ as well as  ${\bf [56,2^+]^2 8}$ and ${\bf [56,2^+]^4 10}$,
is expected to be locally enhanced over the scaling curve for both proton and
neutron targets. Note that to order $q^2$ the ${\bf [56,0^+]}$
and ${\bf [70,1^-]}$ multiplets are sufficient to realise
duality. Formally the analyis can be extended to higher $q^2$
by including correspondingly higher multiplets;
however, the credibility of the non-relativistic harmonic
oscillator may become questionable.
These predictions will be interesting tests of our analysis.

 Inclusion of both magnetic and electric interactions shows that the
duality is non-trivial. Inasmuch as the magnetic terms dominate at large
$Q^2$ in the quark model, duality can be
realised for the dominantly transverse scattering of the deep
inelastic region. For the longitudinal structure function, $F_L$,
duality is again realised, with the breakdown into $\bf 56$ and $\bf 70$
as in Table 3:

\vskip 0.3in Table 3: Relative Longitudinal Production Strengths,
as in Table 2

 $
\begin{array}{lcccccc}
\hline SU(6): &{\bf [56,0^+]^2 8} & {\bf [56,0^+]^4 10} & {\bf
[70,1^-]^2 8} & {\bf [70,1^-]^4 8} & {\bf [70,1^-]^2 10}  & total
 \\
\hline
F_L^p & 1 & 0 & 1 & 0 & 1 &  3 \\
F_L^n & 0 & 0 & 1 & 0 & 1 &  2 \\
\hline
\end{array}
$

\vskip 0.2in

\noindent However, for $F_1(Q^2 \to 0)$ both electric and magnetic multipoles
contribute and interfere with phases determined by the $J^P$ and the
spin-$L_z$ correlations in the various $\bf 56$ and $\bf 70$ states. This
causes dramatic $Q^2$ dependence in polarisation asymmetries \cite{cgk,cg1}
and enables the connection to the Drell-Hearn-Gerasimov sum rule at $Q^2 = 0$.
Thus we predict that {\it Bloom-Gilman duality must fail at $Q^2$ where the
electric and magnetic multipoles have comparable strengths}. Calculations
in simplistic models successfully predicted that this would be at $Q^2 \sim
0.5$ GeV$^2$\cite{cg1}; these results now merit, and are receiving,
 more detailed examination
\cite{N*2000}. The data from Ref. \cite{jlab1} show that even as
low as $Q^2 = 0.5$ (GeV/c)$^2$ the integrated strengths of spectra
at fixed $Q^2$ are within $10 \%$
of the corresponding integrals over the scaling curve. It will be
interesting to verify the predicted breakdown at lower $Q^2$
and also to test if duality in the {\bf magnetic} multipoles
holds all the way to $Q^2 = 0$.

{\bf A Possible Extension to Fragmentation Functions}

Our conclusion that the destructive interference between hadronic
states of different symmetries is a critical feature of duality, can be
applied to semi-inclusive hadroproduction, such as
$\gamma(q) N \to \pi + X$. In the quark parton model in the ideal
valence quark region when $u(x) = 2d(x)$
and at large z,
where $D_u^{\pi^-}/D_u^{\pi^+} \to 0$, it is trivial to obtain
relations such as $F(\gamma p \to \pi^+ + X)/F(\gamma p \to \pi^- + X)
=8$ (= $1/2$ for neutron targets) \cite{inclparton}.
The coherent picture described above can be applied to these
processes. We find that destructive interference leads to
factorisation and to duality, with for example

\begin{equation}
F(\gamma p \to \pi^+ + X)
\equiv \sum_{(W'= {\bf 56,70})}F(\gamma p \to \pi^+ W')
\end{equation}

\noindent where

\begin{equation}
F(\gamma N \to \pi^+ W') =
\sum_{(N^*,N'^*)} F_{(\gamma N \to N^*)}
D_{(N^* \to N'^* \pi)} \sim
\sum_q e_q^2 q(x) D_{q \to \pi}(z)
\end{equation}

\noindent where $D_{q \to \pi}$ is the quark $\to \pi$ fragmentation function,
 $F_{(\gamma N \to N^*)}$ is the $\gamma N \to N^*$ transition form factor,
and $D_{(N^* \to N'^* \pi)}$ is a function representing the decay
$N^* \to N'^* \pi$ where $W'$ is the invariant mass of the final state
$N'^*$.  The breakdown of $F(\gamma N \to \pi W')$
into the individual states in the supermultiplets
for the final $W'$ states is shown in Table 4:

\vskip 0.2in
Table 4:
 $SU(6)$ and $SU(3) \times SU(2)$ Multiplet Contributions to
Inclusive $\pi^{\pm}$ Photproduction

$\begin{array}{lcccc}
\hline
W' & \gamma p \to \pi^+ + W' & \gamma p \to \pi^- + W' &
\gamma n \to \pi^+ + W' & \gamma n \to \pi^- + W' \\
\hline
{\bf 56;8} & 100 & 0 & 0 & 25 \\
{\bf 56;10} & 32 & 24 & 96 & 8 \\
\hline
{\bf 70;^2 8} & 64 & 0 & 0 & 16 \\
{\bf 70;^4 8} & 16 & 0 & 0 & 4 \\
{\bf 70;10} & 4 & 3 & 12 & 1 \\
\hline
TOTAL & 216 & 27 & 108 & 54
\end{array}
$

\vskip 0.2in

\noindent  Note the self-consistency of the results:
 $F(\gamma p \to (\pi^+ + \pi^-))/
F(\gamma n \to (\pi^+ + \pi^-))$ recovers the 3/2 ratio familiar
for the ``total" $F^p/F^n$.
We see also that duality may be obtained at large $Q^2, W^2$  when
$W'$ is integrated over the range up to 1.7 GeV (${\bf 56} + {\bf 70}$).
 We see also
that to a good approximation it may ensue for the ${\bf 56}$ alone when
Born terms and the $\Delta$ are included. To the extent that this is true,
one may expect factorisation and approximate duality at small $Q^2, W'^2 \leq
3$GeV$^2$. A possible example of this was noticed long ago in Ref \cite{bm}
for $\gamma p(n) \to \pi \Delta$ when $0.2 \leq |t| \leq 1 $GeV$^2$.

In  $\gamma N \to \pi + W'$, the above Table strictly applies only to
the imaginary part of the amplitudes. One should in principle
also consider $u-$channel diagrams,
where the $\pi$ is emitted prior to the photoabsorbtion. These diagrams
would give an inversion of the ratios for $\pi^+/\pi^-$.
Ref. \cite{bm} has shown, via fixed $t$ dispersion relations, that
at least for small $Q^2$, the $s$ and $u$-channel resonances tend to
cancel in the real part, whereby the charge ratios are preserved. There
is some indication from data on photoproduction that this is the case
empirically \cite{data,bm}, but the
 dependence of these processes on $Q^2$ needs to
be checked in explicit models and the precision of these ratios tested with
data from CEBAF. We hope to return to these questions elsewhere.

Finally, we comment on a characteristic feature of one particle inclusive
 data: the production of jets.  In this resonance duality approach,
 for any individual partial wave $L$ there is a specified
angular distribution. Upon summing over $L$,
the jet characteristics of the quark parton model arise
as a result of the constructive interference
of spherical harmonics in the near forward and destructive interference
in the backward
hemisphere \cite{IJ}.
As may be anticipated from the uncertainty principle,
after summing over $L$ one expects an angular spread of the jet
given by $\Delta \phi  \sim h/L_{max}$.

In this letter we have outlined the general features of a programme
to understand not only duality, but also to define the
``semi-local" averaging procedures that must be employed to see duality at
low energies.
We shall report elsewhere on detailed predictions that may be
tested in the forthcoming experimental programs for CEBAF
at Jefferson Laboratory. If we are successful in defining these
 averaging procedures, it would have a great impact on our ability to
measure structure functions in kinematic regions that hitherto have been
believed to be inaccessible.

\begin{center}
{\bf Acknowledgements}
\end{center}
\par
We are indebted to W. Melnitchouk, and also to experimentalists at
Jefferson Laboratory for discussions of their data. FEC is indebted to
Jefferson Laboratory for hospitality and support.
This work is supported, in part, by
the European Community Human Mobility Program Eurodafne,
contract NCT98-0169 and by DOE contract DE-AC05-84ER40150
under which the Southeastern Universities Research Associates (SURA)
operates the Thomas Jefferson National Accelerator Facility
(Jefferson Lab).

\end{document}